\documentclass[12pt]{article}
\usepackage{epsfig}
\usepackage{array}
\makeatletter
\def\alt{\mathrel{\mathpalette\gl@align<}}
\def\agt{\mathrel{\mathpalette\gl@align>}}
\def\gl@align#1#2{\lower.6ex\vbox{\baselineskip\z@skip\lineskip\z@
\ialign{$\m@th#1\hfil##\hfil$\crcr#2\crcr\sim\crcr}}}
\makeatother

\begin{document}
\begin{flushright}
{\tt hep-ph/0304118}\\
OSU-HEP-03-5 \\
April, 2003 \\
\end{flushright}
\vspace*{2cm}
\begin{center}
{\baselineskip 25pt
\large{\bf
Unity of elementary particles and forces in higher dimensions
}}

\vspace{1cm}

{\large
Ilia Gogoladze\footnote
{On a leave of absence from: Andronikashvili Institute of Physics, GAS, 380077, Tbilisi, Georgia.\\
email: {\tt ilia@hep.phy.okstate.edu}}, 
Yukihiro Mimura\footnote
{email: {\tt mimura@hep.phy.okstate.edu}}
and
S. Nandi\footnote
{email: {\tt shaown@okstate.edu}}
}
\vspace{.5cm}

{\small {\it Physics Department, Oklahoma State University, \\
             Stillwater, OK 74078}}

\vspace{.5cm}

\vspace{1.5cm}
{\bf Abstract}
\end{center}

We present a model in which elementary particles and forces
are unified in the framework of quantum field theory in higher dimensions.
The particles include gauge bosons, quarks and leptons, as well as the Higgs bosons
and the forces include strong, weak, hypercharge as well as Yukawa interactions.
The model is based on a simple group $SO(16)$ in six dimensions with $N=2$ supersymmetry.
The gauge symmetry, as well as supersymmetry is broken to 
$SU(4) \times SU(2)_L \times SU(2)_R \times U(1)^3$ with $N=1$ supersymmetry
upon compactification in four dimension on a $T^2/Z_6$ orbifold.

\thispagestyle{empty}

\bigskip
\newpage

\addtocounter{page}{-1}

\section{Introduction}
\baselineskip 20pt

Recent topics of the theories in higher dimensions give us a lot of interesting
phenomenological pictures.
We consider that the extra dimensions are compactified in an orbifold.
In an orbifold space,
we can impose transformation properties to the fields 
and the symmetries can be broken 
\cite{Hosotani:1983xw,Scherk:1978ta}.
The gauge symmetry can be also broken down through the orbifold compactification.
Ref.\cite{Kawamura:1999nj} utilizes this orbifold breaking to consider the $SU(5)$ 
grand unified theories,
and solves the doublet-triplet splitting problem by projecting out
the colored Higgs triplets 
using the orbifold transformation properties.
The orbifold grand unified theories are widely applied to build models
in higher dimensions \cite{Hebecker:2001jb}.

The orbifold grand unified theories can give us an interesting situation:
The bulk lagrangian has a larger symmetry, while on the 4 dimensional wall (3-brane) the symmetry is broken.
Such situation provides us a possibility that parameters in the 4D models are related.
For a simple example,
suppose that the bulk symmetry is $SO(10)$
and the bulk symmetry is broken down to Pati-Salam \cite{Pati:1974yy} symmetry, 
$SU(4) \times SU(2)_L \times SU(2)_R$ in 4D.
There is no reason that the three gauge couplings of Pati-Salam symmetry
are related if we only consider the 4D lagrangian.
However, the three gauge couplings can be unified 
due to the larger bulk symmetry 
if we can neglect the brane-localized
gauge kinetic terms by using large volume suppression.
Left-right parity can also originate from large bulk symmetry.
Actually, the Pati-Salam orbifold model is a good example
to consider the orbifold grand unified theories:
Charge quantization is realized naturally, and the electroweak Higgs and
colored Higgs are already split.

Abother attractive motivation to extend the dimensions
is that the variety of particles in Nature
can be understood by means of a geometrical language.
For example, in the original idea by Kaluza-Klein,
the 4 dimensional gauge fields are included in the higher dimensional
metric tensor.
The gauge-Higgs unification \cite{Hosotani:1983xw,Manton:1979kb,Hall:2001tn} is one 
of such attractive ideas in the higher dimensional theories.
Recent progress of the higher dimensional unified theories makes many people
revisit the idea of the gauge-Higgs 
unification \cite{Krasnikov:dt,Hall:2001zb,Burdman:2002se,Haba:2002vc,Gogoladze:2003bb}.
The gauge fields with the extra dimensional components
behave as scalar fields in 4 dimension
and the scalar fields originated from the gauge bosons 
can be Higgs fields which break gauge symmetry.
The masses of such scalar fields are prohibited by gauge invariance,
and in supersymmetric theories
the scalar fields remain massless in the low energy
thanks to the non-renormalizable theorem.
Thus those scalar fields can be good candidates
for the low energy Higgs fields which break electroweak symmetry.

In the higher dimensional supersymmetric theories,
the gauge multiplet, which is the extended supersymmetric multiplet,
contains both vector multiplet and
chiral supermultiplets in 4D $N=1$ language.
Assigning the different transformation property between vector multiplet
and chiral supermultiplets,
we can make vector multiplet massless but chiral supermultiplets heavy in 4D,
which means the extended supersymmetry can be broken down to $N=1$ supersymmetry.
If we also break gauge symmetry through orbifold compactification simultaneously,
the chiral supermultiplets which correspond to the broken generator
can have a zero mode,
which remains massless in the low energy.
Then, we can identify such supermultiplets with the low energy fields.
This is the main idea which we consider in this paper.

The Ref.\cite{Burdman:2002se} emphasizes an interesting possibility
that the gauge and Yukawa coupling constants have the same origin,
if matters (quarks and leptons) are also bulk fields
in the context of gauge-Higgs unification.
The Yukawa interactions arise from the gauge interaction
in the higher dimensional lagrangian,
thus, the gauge and Yukawa coupling constants can be unified in the higher dimensional theory.
This is also realized due to the large bulk symmetry which we mentioned before.
This fact can be a strong motivation to consider the orbifold grand unified theories.
In the Ref.\cite{Burdman:2002se}, the authors considered the 5D theory of $SU(3)_w$ and $SU(6)$
as an example of this scenario.
The gauge-Higgs unification in 5D theory with larger gauge group such as $E_6$, $E_7$ and $E_8$
has also been studied \cite{Haba:2002vc}.
In Ref.\cite{Gogoladze:2003bb}, the authors consider the
gauge-Higgs unification in the $SU(4)_w$ and $SO(12)$, and suggest
that the left-right symmetric breaking of the gauge symmetry gives
an economical realization of the representations of the quarks and
leptons.

Another interesting possibility of unified model in higher dimensions
is that quarks and leptons can be unified in the gauge multiplet.
It is well-known that three (or four) families of quarks and leptons can be contained 
in the adjoint representations in large gauge groups, such as $E_7$ and $E_8$.
The matters in the adjoint representation are always vector-like,
but one can project out the vector-like partner
by $Z_3$ transformation properties \cite{Babu:2002ti}.
This encourages us to consider the interesting possibility that the gauge and matters are 
unified in higher dimensional models.
In other context,
we can consider the origin of the three families as
chiral superfields in gauge multiplet \cite{Watari:2002tf}
since the gauge multiplet in the 6D $N=2$ supersymmetry
contains three $N=1$ chiral superfields in 4D.

In Ref.\cite{Gogoladze:2003ci},
we suggested the possibility of unifying gauge, matter and Higgs fields
in one supersymmetric gauge multiplet in higher dimensions,
as well as the unification of the gauge and Yukawa couplings.
As a simple example,
a 6D $N=2$ supersymmetric $SU(8)$ unified model is constructed.
The gauge symmetry $SU(8)$ is broken down to
Pati-Salam symmetry with two extra $U(1)$'s
in 4D through $T^2/Z_6$ orbifold compactification,
and the theory is reduced to 4D $N=1$ supersymmetric Pati-Salam model.
The electroweak Higgs fields and standard model fermions for the 3rd family
was unified with the gauge bosons in the 6D gauge multiplets.
The 6D bulk gauge interaction produces
Yukawa interactions, which give masses to the quarks and leptons
by Higgs mechanism,
and gives gauge-Yukawa unification.
The numerical agreement of this gauge-Yukawa unification prediction
for all the gauge and the third family Yukawa couplings is good \cite{Gogoladze:2003ci}.

In this paper,
we suggest the extension of our $SU(8)$ unified model
to incorporate two families in the bulk.
We will construct a 6D $N=2$ supersymmetric $SO(16)$ unified model.
In this model,
the gauge symmetry $SO(16)$ is broken down to
Pati-Salam symmetry with three extra $U(1)$'s
in 4D through $T^2/Z_6$ orbifold compactification.
The gauge bosons, Higgs fields and two families of quarks and leptons 
are unified in the 6D $N=2$ gauge multiplets.
Another family is naturally introduced as brane fields to cancel the gauge anomaly.
Thus, this gives the three-family model.
The flavor symmetry can be easily broken down, while
the gauge and Yukawa couplings for 3rd family are unified
in the same way as $SU(8)$ model.

Our paper is organized as follows:
In section 2, we construct our supersymmetric $SO(16)$ model in 6D
with gauge, Higgs and matter unification and show
how orbifold compactification leads to Pati-Salam model in 4D.
Implications of our model are given in section 3.
Section 4 contains our conclusions and discussions.


\section{The Model}


We consider the 6D gauge theory with $N=2$ supersymmetry.
The two extra dimensions are compactified on an orbifold $T^2/Z_n$ \cite{Li:2001dt}.
The $N=2$ supersymmetry in 6D corresponds to $N=4$ supersymmetry
in 4D, thus only the gauge multiplet can be introduced in the bulk.
In terms of 4D $N=1$ language, the gauge multiplet
contains vector multiplet $V(A_\mu,\lambda)$ and three chiral multiplets
$\Sigma$, $\Phi$ and $\Phi^c$ in the adjoint representation
of the gauge group.
The fifth and sixth components of the gauge fields, $A_5$ and $A_6$,
are contained in the lowest component of $\Sigma$, {\it i.e.}
$\Sigma|_{\theta=\bar \theta=0} = (A_6 + i A_5)/\sqrt2$.

The bulk action, written in the 4D $N=1$ language and in the Wess-Zumino gauge,
is given by \cite{Arkani-Hamed:2001tb}
\begin{eqnarray}
S &=& \int d^6x \left\{ {\rm Tr} \left[ \int d^2 \theta \left(
\frac1{4kg^2} W^\alpha W_\alpha + \frac1{kg^2} \left( \Phi^c \partial \Phi
- \sqrt2 \Sigma [\Phi,\Phi^c] \right) \right) + h.c. \right] \right.\nonumber \\
&& + \int d^4 \theta \frac1{kg^2} {\rm Tr} \left[ (\frac1{\sqrt2} \partial^\dagger +\Sigma^\dagger)
e^{-2V} (-\frac1{\sqrt2} \partial + \Sigma) e^{2V} + \frac14 \partial^\dagger e^{-2V} \partial e^{2V} \right]
\nonumber \\
&& + \left. \int d^4 \theta \frac1{kg^2} {\rm Tr} \left[\Phi^\dagger e^{-2V} \Phi e^{2V}
+ \Phi^{c \dagger} e^{-2V} \Phi^c e^{2V} \right] \right\},
\label{6D_action}
\end{eqnarray}
where $k$ is the normalization of the group generators, ${\rm Tr} T^a T^b = k \delta^{ab}$ (we take $k=1/2$),
$\partial$ is defined as $\partial = \partial_5 - i\partial_6$,
and $W_\alpha$ is defined as $W_\alpha = - \frac18 \bar D^2 (e^{-2V} D_\alpha e^{2V})$.
The 6D gauge transformations are
\begin{eqnarray}
e^{2V} &\rightarrow& e^{\Lambda} e^{2V} e^{\Lambda^\dagger}, \quad
\Sigma \rightarrow e^{\Lambda} (\Sigma - \frac1{\sqrt2} \partial ) e^{-\Lambda},
\label{gauge-transform}\\
\Phi   &\rightarrow& e^{\Lambda} \Phi e^{-\Lambda}, \quad
\Phi^c \rightarrow e^{\Lambda} \Phi^c e^{-\Lambda}.
\label{gauge-transform2}
\end{eqnarray}

The $T^2/Z_n$ orbifold is constructed by identifying the complex coordinate $z$ of the extra dimensions
under $Z_n : z \rightarrow \omega z$, where $\omega^n =1 $.
The number $n$ is restricted to be $n=2,3,4,6$.
We can impose the transformation property of the gauge multiplet as
\begin{eqnarray}
V(x^\mu,\omega z, \bar \omega \bar z) &=& {\cal R} V(x^\mu,z,\bar z) ,
\label{transformation1} \\
\Sigma(x^\mu,\omega z, \bar \omega \bar z) &=& \bar \omega \ {\cal R}  \Sigma(x^\mu,z,\bar z),
\label{transformation2} \\
\Phi(x^\mu,\omega z, \bar \omega \bar z) &=& \omega^{l} \ {\cal R} \Phi(x^\mu,z,\bar z),
\label{transformation3}\\
\Phi^c(x^\mu,\omega z, \bar \omega \bar z) &=& \omega^{m} \ {\cal R} \Phi^c(x^\mu,z,\bar z),
\label{transformation4}
\end{eqnarray}
where $\cal R$ acts on the gauge space
satisfying ${\cal R}^n$ to be the identity mapping
since $V(x^\mu,\omega^n z,\bar \omega^n \bar z)$ should be equal to $V(x^\mu,z,\bar z)$.
Non-trivial $\cal R$ breaks the gauge symmetry.
Because of the lagrangian invariance in Eq.(\ref{6D_action})
under the transformations (\ref{transformation1}-\ref{transformation4}),
we have a relation $l+m \equiv 1 $ (mod $n$).
In the case $n>2$, this transformation property breaks $N=4$ supersymmetry down to $N=1$ in 4D.
We will concentrate on $T^2/Z_6$ orbifold in the following
in the same way as in Ref.\cite{Gogoladze:2003ci}.

We consider the $SO(16)$ gauge group.
The adjoint of $SO(16)$ is represented by $16\times 16$ real anti-symmetric matrices.
The gauge twisted mapping ${\cal R}$ is represented as 
${\cal R} V \equiv R V R^T$
where $R$ is a unitary matrix which satisfies that $R^n = \pm I$ ($I$ is the identity matrix).
For example, if we take unitary matrix $R$
\begin{equation}
R = {\rm diag}(1,1,1,1,1,1,1,1) \otimes {\rm diag}(\omega,\bar \omega)
\equiv {\rm diag} (\overbrace{\omega,\cdots,\omega}^8,
\overbrace{\bar \omega,\cdots,\bar \omega}^8),
\label{matrix:1}
\end{equation}
$SO(16)$ is broken down to $SU(8)\times U(1)$.
The adjoint $\mathbf{120}$ is decomposed as
$\mathbf{120} = \mathbf{63}_0 + \mathbf{1}_0 + \mathbf{28}_{-1} + \mathbf{\overline{28}}_1$
under $SU(8) \times U(1)$. The $U(1)$ charges in some normalization are given by
subscript. 
In the following, we denote this $U(1)$ symmetry by $U(1)_3$.
The vector multiplet $V$ is decomposed as 
$V_{\mathbf{120}} = V_{\mathbf{63}} + V_{\mathbf{1}} + V_{\mathbf{28}} + V_{\mathbf{\overline{28}}}$,
and similar for the chiral multiplets $\Sigma$, $\Phi$ and $\Phi^c$.
In the choice of matrix in Eq.(\ref{matrix:1}),
the $Z_6$ charges are assigned as 1 for $V_{\mathbf{63}} + V_{\mathbf{1}}$,
$\omega^2$ for $V_{\mathbf{28}}$ and $\bar \omega^2$ for $V_{\mathbf{\overline{28}}}$.
The $Z_6$ charge assignments for $\Sigma$, $\Phi$ and $\Phi^c$ are obtained by multiplying
$\bar \omega$, $\omega^l$ and $\omega^m$, respectively.

Now we take the matrix $R$ as
\begin{equation}
R = \left(
\begin{array}{c|c}
\omega^{\frac{a}2} R_8 & 0 \\ \hline
0 & \bar \omega^{\frac{a}2} R_8^\dagger
\end{array}
\right)
\label{unitary R}
\end{equation}
where $R_8$ is an $8 \times 8$ unitary matrix which satisfies $R_8^n =I$.
The $Z_6$ transformation property for the vector multiplet
$V$ are easily obtained as
\begin{eqnarray}
V_{\mathbf{63}}(x^\mu,\omega z,\bar \omega \bar z) &=& R_8 V_{\mathbf{63}}(x^\mu,z,\bar z) R_8^\dagger, \\
V_{\mathbf{28}}(x^\mu,\omega z,\bar \omega \bar z) &=& \omega^a R_8 V_{\mathbf{28}}(x^\mu,z,\bar z) R_8, \\
V_{\mathbf{\overline{28}}}(x^\mu,\omega z,\bar \omega \bar z) &=& \bar \omega^a R_8^\dagger V_{\mathbf{\overline{28}}}(x^\mu,z,\bar z) R_8^\dagger, \\
V_{\mathbf{1}}(x^\mu,\omega z,\bar \omega \bar z) &=& V_{\mathbf{1}}(x^\mu,z,\bar z).
\end{eqnarray}
The transformation property for $\Sigma$, $\Phi$ and $\Phi^c$ can be obtained multiplying $\bar \omega$,
$\omega^l$ and $\omega^m$, respectively.
We choose the unitary matrix $R_8$ as
\begin{equation}
R_8 = {\rm diag} (\omega^b,\omega^b,\omega^b,\omega^b,\omega^c,\omega^c,\omega^d,\omega^d).
\label{matrix:2}
\end{equation}
Then,
with this choice ($a$ is odd and 
$b$, $c$, $d$ are different numbers modulo $6$), 
$SO(16)$ breaks down to $SU(4) \times SU(2)_L \times SU(2)_R \times U(1)^3$,
and the 4D theory becomes $N=1$ supersymmetric Pati-Salam model with three extra $U(1)$ symmetries.
%
%
%
%

The $SU(8) \times U(1)_3$ representations $\mathbf{63}_0$, $\mathbf{1}_0$, $\mathbf{28}_{-1}$ and $\mathbf{\overline{28}}_1$
are decomposed under the $SU(4) \times SU(2)_L \times SU(2)_R$ in the following matrix form:
\begin{eqnarray}
\mathbf{63}_0 &\!\!=&\!\! \left(
\begin{array}{ccc}
\mathbf{(15,1,1)}_{0,0,0} & \mathbf{(4,2,1)}_{1,0,0} & \mathbf{(4,1,2)}_{1,2,0} \\
\mathbf{(\bar 4,2,1)}_{-1,0,0} & \mathbf{(1,3,1)}_{0,0,0} & \mathbf{(1,2,2)}_{0,2,0} \\
\mathbf{(\bar 4,1,2)}_{-1,-2,0} & \mathbf{(1,2,2)}_{0,-2,0} & \mathbf{(1,1,3)}_{0,0,0}
\end{array}
\right) + \mathbf{(1,1,1)}_{0,0,0} + \mathbf{(1,1,1)}_{0,0,0}, \\
\mathbf{28}_{-1} &\!\!=&\!\! \left(
\begin{array}{ccc}
\mathbf{(6,1,1)}_{1,1,-1} & \mathbf{(4,2,1)}_{0,1,-1} & \mathbf{(4,1,2)}_{0,-1,-1} \\
 & \mathbf{(1,1,1)}_{-1,1,-1} & \mathbf{(1,2,2)}_{-1,-1,-1} \\
(\mbox{anti-sym})&  & \mathbf{(1,1,1)}_{-1,-3,-1}
\end{array}
\right), \\
\mathbf{\overline{28}}_1 &\!\!=&\!\! \left(
\begin{array}{ccc}
\mathbf{(6,1,1)}_{-1,-1,1} & \mathbf{(\bar 4,2,1)}_{0,-1,1} & \mathbf{(\bar 4,1,2)}_{0,1,1} \\
 & \mathbf{(1,1,1)}_{1,-1,1} & \mathbf{(1,2,2)}_{1,1,1} \\
(\mbox{anti-sym}) &  & \mathbf{(1,1,1)}_{1,3,1}
\end{array}
\right), \qquad
\mathbf{1}_0 = (\mathbf{1,1,1})_{0,0,0}.
\end{eqnarray}
where the subscripts denote the charges under the $U(1)_1 \times U(1)_2 \times U(1)_3$ symmetry.
The $U(1)_1$ and $U(1)_2$ symmetries originate from the $SU(8)$ generators:
${\rm diag}(1,1,1,1,-1,-1,-1,-1)/2$ for $U(1)_1$ and ${\rm diag}(1,1,1,1,1,1,-3,-3)/2$ for $U(1)_2$.
The $Z_6$ transformation property for these decomposed representations of the vector multiplet $V$
are easily calculated as
\begin{eqnarray}
V_{\mathbf{63}} &\!\!:&\!\! \left(
\begin{array}{ccc}
1 & \omega^{b-c} & \omega^{b-d} \\
\omega^{c-b} & 1 & \omega^{c-d} \\
\omega^{d-b} & \omega^{d-c} & 1
\end{array}
\right) + (1) + (1), \qquad
V_{\mathbf{1}} : (1), 
\label{z6-63} \\
V_{\mathbf{28}} &\!\!:&\!\! \left(
\begin{array}{ccc}
\omega^{a+2b} & \omega^{a+b+c} & \omega^{a+b+d} \\
 & \omega^{a+2c} & \omega^{a+c+d} \\
 &  & \omega^{a+2d}
\end{array}
\right), \qquad
V_{\mathbf{\overline{28}}} : \left(
\begin{array}{ccc}
\bar \omega^{a+2b} & \bar \omega^{a+b+c} & \bar \omega^{a+b+d} \\
 & \bar \omega^{a+2c} & \bar \omega^{a+c+d} \\
 &  & \bar \omega^{a+2d}
\label{z6-28}
\end{array}
\right).
\end{eqnarray}
All the exponents of $\omega$ should be non-zero modulo 6.
The $Z_6$ assignments for $\Sigma$, $\Phi$ and $\Phi^c$ are obtained by multiplying
$\bar \omega$, $\omega^l$ and $\omega^m$, respectively.

We choose the matrix $R_8$ as
\begin{equation}
R_8 = {\rm diag} (1,1,1,1,\omega^5, \omega^5, \omega^2, \omega^2) 
\end{equation}
to pick up one chiral family from $\mathbf{63}$.
Then we find the $Z_6$ transformation property of the vector multiplet $V$
from Eqs.(\ref{z6-63}) and (\ref{z6-28}) as
\begin{eqnarray}
V_{\mathbf{63}} &\!\!:&\!\! \left(
\begin{array}{ccc}
1 & \omega & \omega^4 \\
\omega^5 & 1 & \omega^3 \\
\omega^2 & \omega^3 & 1
\end{array}
\right) + (1) + (1), \qquad
V_{\mathbf{1}} : (1), 
\\
V_{\mathbf{28}} &\!\!:&\!\! \omega^a \left(
\begin{array}{ccc}
1 & \omega^5 & \omega^2 \\
 & \omega^4 & \omega \\
 &  & \omega^4
\end{array}
\right), \qquad
V_{\mathbf{\overline{28}}} : \bar \omega^a \left(
\begin{array}{ccc}
1 & \omega & \omega^4 \\
 & \omega^2 & \omega^5 \\
 &  & \omega^2
\end{array}
\right).
\end{eqnarray}
The number $a$ is chosen as $a=3$ to make all the $V_{\mathbf{28}}$ and $V_{\mathbf{\overline{28}}}$ massive,
and then
\begin{equation}
V_{\mathbf{28}} : \left(
\begin{array}{ccc}
\omega^3 & \omega^2 & \omega^5 \\
 & \omega & \omega^4 \\
 &  & \omega
\end{array}
\right), \qquad
V_{\mathbf{\overline{28}}} :  \left(
\begin{array}{ccc}
\omega^3 & \omega^4 & \omega \\
 & \omega^5 & \omega^2 \\
 &  & \omega^5
\end{array}
\right).
\end{equation}
With this choice, the gauge symmetry $SO(16)$ is broken down to $SU(4) \times SU(2)_L \times SU(2)_R 
\times U(1)^3$.
We can extract zero-modes in the chiral superfields $\Sigma$, $\Phi$ and $\Phi^c$
through the transformation property (\ref{transformation2}-\ref{transformation4})
with $(l,m)=(4,3)$.
The zero-modes are listed in the Table 1.
The $L_i$ and $\bar R_i$ include left- and right-handed quarks and leptons 
and $C_i$ includes vector-like colored Higgs.
The model thus includes two chiral families in the bulk and three electroweak
bidoublets $H_i$.

\begin{table}
\begin{center}
\begin{tabular}{|c||c|c|c|}
\hline
& $\mathbf{63}$ & $\mathbf{28}$ & $\mathbf{\overline{28}}$ \\ \hline \hline
$\Sigma$ & $L_3: \mathbf{(4,2,1)}_{1,0,0}$ &
\begin{tabular}{ll} $S_1:$ & $\mathbf{(1,1,1)}_{-1,1,-1}$ \\
                            $S_2:$ &$\mathbf{(1,1,1)}_{-1,-3,-1}$
\end{tabular}
& $\bar R_2: \mathbf{(\bar 4,1,2)}_{0,1,1}$ \\  \hline
$\Phi$ &
$\bar R_3:\mathbf{(\bar 4,1,2)}_{-1,-2,0}$ &
$L_2:\mathbf{(4,2,1)}_{0,1,-1}$ &
$H_3:\mathbf{(1,2,2)}_{1,1,1}$ \\ 
\hline
$\Phi^c$ &
\begin{tabular}{ll}
$H_1:$&$\mathbf{(1,2,2)}_{0,2,0}$ \\
$H_2:$&$\mathbf{(1,2,2)}_{0,-2,0}$
\end{tabular} &
$C_1: \mathbf{(6,1,1)}_{1,1,-1}$ & $C_2: \mathbf{(6,1,1)}_{-1,-1,1}$
\\ \hline
\end{tabular}
\end{center}
\caption{List of the zero-modes in chiral multiplets.}
\end{table}

Since the three $N=1$ chiral multiplets $\Sigma$, $\Phi$ and $\Phi^c$ are in the gauge multiplets,
those fields have gauge interactions with each other in 6D.
The trilinear interaction term from Eq.(\ref{6D_action})
\begin{equation}
S = \int d^6 x \left[\int d^2 \theta \frac1{kg^2} {\rm Tr} \left(- \sqrt2 \Sigma [\Phi,\Phi^c]\right)
+ h.c. \right]
\label{trilinear term}
\end{equation}
includes bulk superpotential for the zero modes,
\begin{equation}
S= \int d^6x \int d^2 \theta \ y_6 \left(
L_3 H_1 \bar R_3 + L_2 H_2 \bar R_2
+ (H_1 S_2 - H_2 S_1) H_3
+ \bar R_3 C_1 \bar R_2 + L_3 C_2 L_2 \right) + h.c.,
\label{bulk-Yukawa}
\end{equation}
which includes Yukawa couplings.

Taking into account the normalization factors of the wave functions in the kinetic term,
we find that the six dimensional Yukawa coupling is equal to six dimensional gauge coupling,
$y_6 = g_6$.
The corresponding four dimensional couplings are derived as the coordinates of extra dimensions
are integrated out in the action.
Thus, the four dimensional Yukawa and gauge coupling can be the same dimensionless
number
if the following conditions are satisfied \cite{Gogoladze:2003ci}:
1) The brane-localized gauge and Yukawa interactions 
and their threshold corrections can be negligible.
2) The zero modes of fermions are not localized
at different points on the orbifold.
3) The four dimensional fields are not largely mixed with other brane-localized fields.

\section{Implications of the Model}

We first discuss the implications of the bulk interaction in Eq. (\ref{bulk-Yukawa}).
Suppose that the vacuum expectation values are given to $S_1$ and $S_2$ which are the singlet
under Pati-Salam symmetry.
Then one linear combination of bidoublet Higgses and $H_3$ become heavy, and the following linear combination
remains light:
\begin{equation}
H = \frac1{\sqrt{S_1^2 + S_2^2}} (H_1 S_1 + H_2 S_2).
\end{equation}
The Yukawa coupling terms are rewritten by using this light bidoublet Higgs as
\begin{equation}
L_3 H_1 \bar R_3 + L_2 H_2 \bar R_2 \rightarrow 
\frac1{\sqrt{S_1^2 + S_2^2}} ( S_1 L_3 H \bar R_3 + S_2 L_2 H \bar R_2).
\label{flavor-violation}
\end{equation}
The bulk superpotential term has a $Z_2$ flavor symmetry
such as
\begin{equation}
L_3 \leftrightarrow L_2, \quad \bar R_3 \leftrightarrow \bar R_2, \quad
H_1 \leftrightarrow H_2, \quad S_1 \leftrightarrow S_2, \quad H_3 \leftrightarrow -H_3,
\quad C_1 \leftrightarrow -C_1, \quad C_2 \leftrightarrow -C_2.
\end{equation}
The vacuum expectation values of $S_1$ and $S_2$ can break this $Z_2$ symmetry.
Assuming $\langle S_1 \rangle \gg \langle S_2 \rangle$, we obtain the fermion mass hierarchy
between the 3rd and 2nd family,
supposing that $L_3$ and $\bar R_3$ are for the third generation
and $L_2$ and $\bar R_2$ are for the second generation.
Of course, this is just a toy structure,
since we still have a wrong relation, $m_c/m_t = m_s/m_b = m_\mu/m_\tau$.
We don't have flavor mixing in the bulk superpotential either.
These problems can be solved by 
introducing brane-localized interaction.
The vacuum expectation values of $SU(2)_R$ triplet and $SU(4)$ adjoint Higgs
can break the wrong mass relation in the similar way as in the usual Pati-Salam model.

We comment on the colored Higgs $C_1$ and $C_2$. 
The colored Higgs can get mass as a brane-localized term 
such as $m C_1 C_2$.
The mass term does not violate the baryon (lepton) number conservation.
The mass terms such as $m C_1^2$ or $m C_2^2$ violate the conservation,
but such mass terms are forbidden by extra $U(1)$ symmetry.

Since we project out the vector-like partners by $Z_6$,
the remaining fermion components in Table 1
give rise to gauge anomaly for the two linear combinations of three extra $U(1)$ symmetries.
Green-Schwarz mechanism \cite{Green:sg} can be used to cancel out for only one linear combination.
Thus we have to introduce other brane fields which are non-singlets
under Pati-Salam symmetry to cancel the anomaly.
This can be interpreted as the origin of the 1st family.
For example,
if we introduce brane fields such as
\begin{equation}
L_1 : \mathbf{(4,2,1)}_{-1,-1,1}, \quad \bar R_1 : \mathbf{(\bar 4,1,2)}_{1,1,-1},
\quad H_4 : \mathbf{(1,2,2)}_{-1,-1,-1},
\end{equation}
the anomaly such as $SU(4)^2 \times U(1)$, $SU(2)_L^2 \times U(1)$ and $SU(2)_R^2 \times U(1)$ 
are cancelled out.
Then, introducing appropriate singlet under Pati-Salam symmetry with non-zero extra $U(1)$ charges,
we can obtain anomaly free particle contents.
If we adopt the Green-Schwarz mechanism, we can 
make it anomaly free by introducing $L_1$ and $\bar R_1$ with appropriate
$U(1)$ charges without introducing $H_4$.
In any case, this model contains three families naturally.

We identify the two families originating from gauge multiplet with the 3rd and 2nd (or 1st) family,
and 1st (or 2nd) family is brane-localized fields at 3-brane fixed point.
Then the Yukawa couplings for 3rd family, $y_t$, $y_b$ and $y_\tau$,
are unified to the gauge couplings at compactification scale,
if we neglect a small correction in Eq.(\ref{flavor-violation}),
and the correction coming from brane-localized interaction by assuming
large volume suppression.
The hierarchy of Yukawa couplings for 2nd (or 1st) family
is derived by assuming $\langle S_1 \rangle \gg \langle S_2 \rangle$ as we mentioned before,
and the Yukawa couplings for 1st (or 2nd) family
are naturally small
since their values are suppressed by volume factor of the extra dimensions.

We assume that the compactification scale from 6D to 4D
is the same scale where $SU(4) \times SU(2)_L \times SU(2)_R
\times U(1)^2$ gauge symmetry are broken to Standard Model one,
choosing appropriate Higgs superfields.
So below the compactification scale, we have usual MSSM particle content with
gauge-Yukawa unification condition for particles from third
family
\begin{equation}
\alpha_1=\alpha_2=\alpha_3=\alpha_t=\alpha_b=\alpha_{\tau}
\end{equation}
where $\alpha_1$, $\alpha_2$ and $\alpha_3$ corresponds
hypercharge (with proper normalization), weak and strong
interaction couplings and $\alpha_t$, $\alpha_b$, $\alpha_{\tau}$ are the top,
bottom and tau Yukawa coupling respectively. 
We use the
notation $y^2_{t,b,\tau}/4\pi\equiv\alpha_{t,b,\tau}$. Note that
we neglect brane-localized gauge kinetic terms.


Due to a crucial reduction of the number of the fundamental
parameters from the gauge-Yukawa coupling unification, we are lead immediately
to a series of the very distinctive predictions (in absence of any
large supersymmetric threshold corrections). Using the values of the
electroweak parameters $\sin^2\theta_w=0.2311\pm0.0001$ and
$\alpha_{EM}=127.92\pm0.02$ at $M_Z$ scale \cite{PDG},
we can determine the unification scale and unified coupling constant.
Then, evolving the remaining couplings from the unification scale
to the low energy, we predict\footnote{The numerical calculation of gauge-Yukawa
unification in a 4D model is demonstrated in the Ref.\cite{Chkareuli:1998wi}.} \cite{Gogoladze:2003ci}
\begin{eqnarray}
\alpha _{3}(M_{Z}) =0.123,~~~m_{t}=178~{\rm GeV},~~~
\frac{m_{b}}{m_{\tau }}(M_{Z}) =1.77,~~~\tan\beta =51. \label{8}
\end{eqnarray}
These are in good agreement with experimental data \cite{PDG}
except the small discrepancy for $\alpha_3$ (world average
value is $\alpha_3=0.117\pm0.002$ \cite{PDG}).
The small discrepancy for $\alpha_3$ can be easily
improved if we consider unification scale threshold of 
$SU(4)$ sextet, $C_1$ and $C_2$.

We have made a choice that $L_3$ and $\bar R_2$ are in the chiral multiplet $\Sigma$
and $\bar R_3$ and $L_2$ are in the $\Phi$.
Since the gauge transformations of $\Sigma$ and $\Phi$
in Eqs.(\ref{gauge-transform}-\ref{gauge-transform2})
are different,
brane-localized 4D lagrangian is not left-right symmetric.
The electroweak Higgs fields, $H_1$ and $H_2$,
are in the chiral multiplets $\Phi^c$
and its gauge transformation is linear,
and thus we can introduce the brane-localized Yukawa coupling terms
which give the CKM mixing angles.

We can also introduce the brane-localized right-handed neutrino mass terms for $\bar R_3$
with the brane fields $\mathbf{(4,1,2)}$,
and the neutrino mass becomes small through the seesaw mechanism.
However, $\bar R_2$ is in the $\Sigma$ whose gauge transformation is non-homogenous,
and thus we cannot introduce the brane-localized Majorana mass term for $\bar R_2$.
Therefore, the Majorana mass term for $\bar R_2$ should arise from bulk interaction.
One might think that we can use Wilson-loop operator to make gauge transformation homogeneous
as used in the Ref.\cite{Hall:2001zb}.
Actually, in 5D models with $S^1$ compactification, we can use the Wilson-loop operator,
${\cal P} \exp(i\oint \Sigma dy)$, $\Sigma |_{\theta=0} = (\sigma + i A_5)/\sqrt2$.
However, since its simple extension breaks chirality
in 6D models with $T^2$ compactification, $\Sigma |_{\theta=0} = (A_6 + i A_5)/\sqrt2$,
we cannot use the Wilson-loop operator.
Thus, we will use higher order bulk interactions.
Since $A \equiv \Phi^c \partial \Phi - \sqrt2 \Phi^c [\Sigma, \Phi]$ is gauge covariant,
$A \rightarrow e^{\Lambda} A e^{-\Lambda}$ under gauge transformation,
$({\rm Tr} A)^2 + {\rm Tr} A^2$ can be gauge invariant bulk interactions.
Such higher order bulk interactions include Majorana mass terms of $\bar R_2$,
if we assume that $\mathbf{(4,1,2)}$ component in $\Phi$ and $\Phi^c$ get vacuum expectation values.
The vacuum expectation values of $\mathbf{(4,1,2)}$ component 
also break Pati-Salam symmetry down to Standard Model one.
The vacua should satisfy the following $F$- and $D$-flat conditions,
\begin{eqnarray}
-F^*_\Sigma &=& - \sqrt2 [\Phi,\Phi^c] =0, \\
-F^*_\Phi &=& -\partial \Phi^c - \sqrt2 [\Phi^c,\Sigma] = 0, \\
-F^*_{\Phi^c} &=& \partial \Phi - \sqrt2 [\Sigma, \Phi] = 0, \\
D &=& \frac12 (\bar \partial \Sigma + \partial \bar \Sigma + [\Sigma, \bar \Sigma]) 
+ \frac12 ( [\Phi,\bar \Phi] + [\Phi^c, \bar \Phi^c])= 0.
\end{eqnarray}
If the vacua of $\Sigma$, $\Phi$ and $\Phi^c$ are commutative and holomorphic function of $z$ or $\bar z$,
the $F$- and $D$- flat conditions are satisfied.
Because of double periodicity, the vacua should be elliptic functions.
We obtain the 4D Majorana mass terms by integrating out with respect to extra dimensions on the
fundamental area of $T^2/Z_6$.

\section{Conclusion and Discussion}

The idea that the Higgs and/or the matter are unified in
the higher dimensional supersymmetric gauge multiplet
leads us naturally to the world in which the standard gauge group in 4D is unified
in grand unified gauge group in higher dimensions.

We have considered the 6D $N=2$ supersymmetric $SO(16)$ gauge theory
in the orbifold $T^2/Z_6$.
The gauge group $SO(16)$ is broken down
by $Z_6$ transformation property,
and the model is reduced to $N=1$ supersymmetric Pati-Salam model
$SU(4) \times SU(2)_L \times SU(2)_R$
with three extra $U(1)$ symmetries in 4D.

In this model,
bidoublet Higgs fields and two families of quarks and leptons
are unified in the gauge sector in the 6D $N=2$ gauge supermultiplet.
The vector-like partners of the quarks and leptons
are projected out by $Z_6$ transformation property.
The explicit mass terms of the Higgs fields
are prohibited by the gauge invariance, and the $N=1$ supersymmetry
preserve the masses of the Higgs fields to the electroweak scale.
Furthermore, because of the unification of Higgs and matter in gauge multiplet,
the 4D Yukawa interactions, which we need in order to give masses
to the fermions by Higgs mechanism, arise from gauge interaction
in 6D lagrangian.
This is the most interesting feature of this model.

One of the families in gauge multiplet
can be identified to the 3rd family,
and then we meet an attractive possibility
that 3rd family Yukawa couplings can be unified to the gauge coupling
at grand unified scale.
The numerical results of the renormalization group flow of the gauge as well as the Yukawa couplings
give good agreement
with low energy data for the top quark mass and bottom-tau mass ratio
with large $\tan \beta$.
Since two linear combinations of extra $U(1)$ symmetries give gauge anomalies,
we have to introduce additional fields to cancel them.
The additional fields can be identified with another family.
So this model has three families naturally.
Though 
the bulk interaction gives wrong relations for the mass ratios of quarks and leptons
in the same way as in the naive Pati-Salam model,
the structure of bulk interaction is interesting and is easily corrected.

Since $SO(16)$ is one of the regular maximal subgroup of $E_8$,
the $E_8$ gauge theory is easily reduced to our $SO(16)$ model
considering the periodic boundary conditions:
$V(x^\mu, z + 2\pi R,\bar z + 2\pi R) = {\cal T} V(x^\mu,z,\bar z)$
and using the same conditions for chiral multiplets $\Sigma$, $\Phi$ and $\Phi^c$.
The $E_8$ adjoint $\mathbf{248}$ is decomposed to $\mathbf{120} + \mathbf{128}$ 
under $SO(16)$.
Assigning $T=+$ for $\mathbf{120}$ and $T=-$ for $\mathbf{128}$,
the spinor representation $\mathbf{128}$'s are projected out,
and $E_8$ symmetry is broken down to $SO(16)$.
In some scenarios in superstring theory,
quarks and leptons are included in the spinor representation $\mathbf{128}$,
while in our model matters and Higgses are included in $\mathbf{120}$,
and its implications need further investigation.

\section*{Acknowledgments}

We thank K.S. Babu, R.N. Mohapatra, H.P. Nilles and S. Raby for useful
discussions.
This work was supported in part by US DOE Grants \# DE-FG030-98ER-41076
and DE-FG-02-01ER-45684.


\begin{thebibliography}{99}
%
%
%
%
%
%
%
\bibitem{Hosotani:1983xw}
Y.~Hosotani,
Phys.\ Lett.\ B {\bf 126}, 309 (1983);
%
Phys.\ Lett.\ B {\bf 129}, 193 (1983);
%
Phys.\ Rev.\ D {\bf 29}, 731 (1984);
%
Annals Phys.\  {\bf 190}, 233 (1989).

%
%
%
%
%
%
%
%

\bibitem{Scherk:1978ta}
J.~Scherk and J.~H.~Schwarz,
Phys.\ Lett.\ B {\bf 82}, 60 (1979);
%
Nucl.\ Phys.\ B {\bf 153}, 61 (1979);

E.~Witten,
Nucl.\ Phys.\ B {\bf 258}, 75 (1985);

P.~Candelas, G.~T.~Horowitz, A.~Strominger and E.~Witten,
Nucl.\ Phys.\ B {\bf 258}, 46 (1985);

%
L.~Dixon, J.~Harvey, C.~Vafa and E.~Witten,
%
Nucl.\ Phys.\ B {\bf 261}, 651 (1985).



%
%
%


\bibitem{Kawamura:1999nj}
Y.~Kawamura,
Prog.\ Theor.\ Phys.\  {\bf 103}, 613 (2000)
[hep-ph/9902423];
%
Prog.\ Theor.\ Phys.\  {\bf 105}, 999 (2001)
[hep-ph/0012125];
%
Prog.\ Theor.\ Phys.\  {\bf 105}, 691 (2001)
[hep-ph/0012352].


%
%
%





%
\bibitem{Hebecker:2001jb}
{\it see for example}

G.~Altarelli and F.~Feruglio,
%
Phys.\ Lett.\  B {\bf 511}, 257 (2001)
[hep-ph/0102301];
%

A.~B.~Kobakhidze,
Phys.\ Lett.\ B {\bf 514}, 131 (2001)
[hep-ph/0102323];
%

L.~J.~Hall and Y.~Nomura,
Phys.\ Rev.\ D {\bf 64}, 055003 (2001)
[hep-ph/0103125];
%
Phys.\ Rev.\ D {\bf 65}, 125012 (2002)
[hep-ph/0111068];
%
Phys.\ Rev.\ D {\bf 66}, 075004 (2002)
[hep-ph/0205067];

Y.~Nomura, D.~R.~Smith and N.~Weiner,
Nucl.\ Phys.\ B {\bf 613}, 147 (2001)
[hep-ph/0104041];

A.~Hebecker and J.~March-Russel,
%
Nucl.\ Phys.\ B{\bf 613}, 3 (2001)
[hep-ph/0106166];
%
Nucl.\ Phys.\ B {\bf 625}, 128 (2002)
[hep-ph/0107039];

T.~j.~Li,
Phys.\ Lett.\ B {\bf 520}, 377 (2001)
[hep-th/0107136];
Nucl.\ Phys.\ B {\bf 619}, 75 (2001)
[hep-ph/0108120];

N.~Haba, Y.~Shimizu, T.~Suzuki and K.~Ukai,
Prog.\ Theor.\ Phys.\  {\bf 107}, 151 (2002)
[hep-ph/0107190];

T.~Asaka, W.~Buchmuller and L.~Covi,
Phys.\ Lett.\ B {\bf 523}, 199 (2001)
[hep-ph/0108021];



L.~J.~Hall, Y.~Nomura, T.~Okui and D.~R.~Smith,
Phys.\ Rev.\ D {\bf 65}, 035008 (2002)
[hep-ph/0108071];

R.~Dermisek and A.~Mafi,
Phys.\ Rev.\ D {\bf 65}, 055002 (2002)
[hep-ph/0108139];



F.~Paccetti Correia, M.~G.~Schmidt and Z.~Tavartkiladze,
Nucl.\ Phys.\ B {\bf 649}, 39 (2003)
[hep-ph/0204080];

%










H.~D.~Kim and S.~Raby,
JHEP {\bf 0301}, 056 (2003)
[hep-ph/0212348].


%
%
%
%
%


\bibitem{Pati:1974yy}
J.~C.~Pati and A.~Salam,
Phys.\ Rev.\ D {\bf 10}, 275 (1974).


%
%
%



\bibitem{Manton:1979kb}
N.~S.~Manton,
Nucl.\ Phys.\ B {\bf 158}, 141 (1979);

D.~B.~Fairlie,
J.\ Phys.\ G {\bf 5}, L55 (1979);
%
Phys.\ Lett.\ B {\bf 82}, 97 (1979);

P.~Forgacs and N.~S.~Manton,
Commun.\ Math.\ Phys.\  {\bf 72}, 15 (1980);

G.~Chapline and R.~Slansky,
Nucl.\ Phys.\ B {\bf 209}, 461 (1982);

D.~Kapetanakis and G.~Zoupanos,
Phys.\ Rept.\  {\bf 219}, 1 (1992).
%
%


%
%
%
%
%
%

\bibitem{Hall:2001tn}
L.~J.~Hall, H.~Murayama and Y.~Nomura,
Nucl.\ Phys.\ B {\bf 645}, 85 (2002)
[hep-th/0107245];

R.~Dermisek, S.~Raby and S.~Nandi,
Nucl.\ Phys.\ B {\bf 641}, 327 (2002)
[hep-th/0205122].



%
%






\bibitem{Krasnikov:dt}
I.~Antoniadis,
Phys.\ Lett.\ B {\bf 246}, 377 (1990);

N.~V.~Krasnikov,
Phys.\ Lett.\ B {\bf 273}, 246 (1991);

I.~Antoniadis and K.~Benakli,
Phys.\ Lett.\ B {\bf 326}, 69 (1994)
[hep-th/9310151];

H.~Hatanaka, T.~Inami and C.~S.~Lim,
Mod.\ Phys.\ Lett.\ A {\bf 13}, 2601 (1998)
[hep-th/9805067];

G.~R.~Dvali, S.~Randjbar-Daemi and R.~Tabbash,
Phys.\ Rev.\ D {\bf 65}, 064021 (2002)
[hep-ph/0102307];

N.~Arkani-Hamed, A.~G.~Cohen and H.~Georgi,
Phys.\ Lett.\ B {\bf 513}, 232 (2001)
[hep-ph/0105239];

I.~Antoniadis, K.~Benakli and M.~Quiros,
New J.\ Phys.\  {\bf 3}, 20 (2001)
[hep-th/0108005];

C.~Csaki, C.~Grojean and H.~Murayama,
hep-ph/0210133.


%
%

\bibitem{Hall:2001zb}
L.~J.~Hall, Y.~Nomura and D.~R.~Smith,
Nucl.\ Phys.\ B {\bf 639}, 307 (2002)
[hep-ph/0107331].

%
%

\bibitem{Burdman:2002se}
G.~Burdman and Y.~Nomura,
Nucl.\ Phys.\ B {\bf 656}, 3 (2003)
[hep-ph/0210257].

%
%
%
%
%


\bibitem{Haba:2002vc}
N.~Haba and Y.~Shimizu,
hep-ph/0212166.


%
%

\bibitem{Gogoladze:2003bb}
Q.~Shafi and Z.~Tavartkiladze,
Phys.\ Rev.\ D {\bf 66}, 115002 (2002);

I.~Gogoladze, Y.~Mimura and S.~Nandi,
hep-ph/0301014, to appear in Phys.\ Lett.\ B. 



%
%



\bibitem{Babu:2002ti}
K.~S.~Babu, S.~M.~Barr and B.~s.~Kyae,
Phys.\ Rev.\ D {\bf 65}, 115008 (2002)
[hep-ph/0202178].

%
%
%
%
%
%

\bibitem{Watari:2002tf}
T.~Watari and T.~Yanagida,
Phys.\ Lett.\ B {\bf 532}, 252 (2002)
[hep-ph/0201086];
%
Phys.\ Lett.\ B {\bf 544}, 167 (2002)
[hep-ph/0205090].

%
%
%
%
%
%

\bibitem{Gogoladze:2003ci}
I.~Gogoladze, Y.~Mimura and S.~Nandi,
hep-ph/0302176, to appear in Phys.\ Lett.\ B.

%
%
%
%
%
%
%
%


\bibitem{Li:2001dt}
T.~j.~Li,
Eur.\ Phys.\ J.\ C {\bf 24}, 595 (2002)
[hep-th/0110065];
Nucl.\ Phys.\ B {\bf 633}, 83 (2002)
[hep-th/0112255];


C.~S.~Huang, J.~Jiang, T.~j.~Li and W.~Liao,
Phys.\ Lett.\ B {\bf 530}, 218 (2002)
[hep-th/0112046].







%
%
%
%
%
%

\bibitem{Arkani-Hamed:2001tb}
N.~Arkani-Hamed, T.~Gregoire and J.~Wacker,
JHEP {\bf 0203}, 055 (2002)
[hep-th/0101233].


%
%
%
%
%
%

\bibitem{Green:sg}
M.~B.~Green and J.~H.~Schwarz,
Phys.\ Lett.\ B {\bf 149}, 117 (1984).

%
%
%
%
%




%
%
%
%



%
%
%
%
%

\bibitem{Chkareuli:1998wi}
J.~L.~Chkareuli and I.~G.~Gogoladze,
Phys.\ Rev.\ D {\bf 58}, 055011 (1998)
[hep-ph/9803335].





\bibitem{PDG} Particle Data Group, K. Hagiwara et. al, Phys. Rev.
D {66}, (2002) 010001-173.

\end{thebibliography}
\end{document}